\documentclass[11pt]{article}
\usepackage{DGfest,epsfig}
\usepackage{times}
\usepackage{graphicx}

\def\Journal#1#2#3#4{{#1} {\bf #2}, #3 (#4)}

\def\NPB{{\em Nucl. Phys.} B}
\def\PLB{{\em Phys. Lett.}  B}

\def\PRD{{\em Phys. Rev.} D}

\def\SJETP{{\em Sov. Phys. JETP}}
\def\JETPL{{\em JETP Lett.}}
\def\SJNP{{\em Sov. J. Nucl. Phys.}}
\def\PAN{{\em Phys. Atom. Nucl.}}
\def\EPJC{{\em Eur. Phys. J.} C}
\def\NPPS{{\em Nucl. Phys.} [Proc. Suppl.]}
\def\JHEP{{\em JHEP}}

\def\be{\begin{equation}}
\def\ee{\end{equation}}
\def\bea{\begin{eqnarray}}
\def\eea{\end{eqnarray}}

\newcommand{\beq}[1]{
\begin{equation}\label{#1}}
\newcommand{\eeq}{\end{equation}}
\newcommand{\bear}[1]{
\begin{eqnarray}\label{#1}}
\newcommand{\eear}{\end{eqnarray}}

\begin{document}
\baselineskip 11.5pt
\title{QCD IN THE REGGE LIMIT: FROM GLUON REGGEIZATION TO PHYSICAL 
AMPLITUDES}

\author{A. PAPA}

\address{Dipartimento di Fisica, Universit\`a della Calabria,\\
Istituto Nazionale di Fisica Nucleare, Gruppo collegato di Cosenza,\\
Arcavacata di Rende, I-87036 Cosenza, Italy}

%

\maketitle\abstracts{
This paper is a brief survey of the Balitskii-Fadin-Kuraev-Lipatov (BFKL)
approach for the description of hard or semi-hard processes in the so-called
Regge limit of perturbative QCD. The starting point is a fundamental
property of perturbative QCD, the gluon Reggeization.
This property, combined with $s$-channel unitarity, allows to predict the 
growth in energy of the amplitude of hard or semi-hard processes 
with exchange of vacuum quantum numbers in the $t$-channel. When also the 
so-called impact factors of the colliding particles are known, then not 
only the behavior with the energy, but the complete amplitude can be determined. 
This was recently done, for the first time with next-to-leading order accuracy
and for a process with colorless external particles, in the case of the 
electroproduction of two light vector mesons.}

\section{Introduction}

The BFKL equation~\cite{BFKL} became very popular in the last years due to
the experimental results on deep inelastic scattering of electrons on 
protons obtained at the HERA collider. These results show a power-growth 
of the gluon density in the proton when the fraction of the proton momentum
carried by the gluon (i.e. the $x$ Bjorken variable) decreases.
 
Together with the DGLAP evolution equation~\cite{DGLAP}, the BFKL equation
can be used for the description of structure functions for the deep inelastic 
electron-proton scattering at small values of the $x$ variable.
It applies in general to all processes where a ``hard'' 
scale exists (the ``hardness'' can be supplied either by a large virtuality 
or by the mass of heavy quarks) which allows the use of perturbation theory. 
It is an iterative integral equation for the determination
of the Green's function for the elastic diffusion of two Reggeized gluons 
(Section~2). The knowledge of this Green's function is enough for the determination
of the growth in energy of the amplitude of hard or semi-hard processes and, for the
case of deep-inelastic electron-proton scattering, for the growth in $x$ of the
gluon density in the proton for decreasing $x$. The kernel of this equation
was first derived in the leading logarithmic approximation (LLA), which means 
resummation of all terms of the type $(\alpha_{s}\ln s)^{n}$, where $\alpha_{s}$ 
is the QCD coupling constant and $s$ is the square of the c.m.s. energy. 
In this approximation total cross sections are predicted to grow at large $s$
with a power of the center-of-mass energy larger than for ``soft'' hadronic 
processes (Section~3).

Unfortunately, in this approximation neither the scale of $s$ nor the argument of 
the running coupling constant $\alpha_{s}$ can be fixed. So, in order to do accurate
theoretical predictions, it was necessary to calculate the radiative corrections 
to the LLA (Section~4). These corrections turned out to be large and with negative
sign with respect to the leading order and this has raised an ongoing debate on the 
reliability of perturbation theory in this context.
Many attempts of improvement have been suggested, but a definite 
conclusion is still lacking, mainly because of the difficulty to translate any
recipe designed to work for the BFKL Green's function, which is an unphysical 
object, into a definite prediction, checkable by experiments.

The information which is needed together with the BFKL Green's function in order
to build a physical amplitude is given by the so-called ``impact factors''
of the colliding particles. At the next-to-leading order impact factors were
calculated first for the case of colliding partons~\cite{FFKP00,CR00}.
Among the impact factors for transitions between colorless objects, the most 
important one from the phenomenological point of view is certainly the 
impact factor for the virtual photon to virtual photon transition, i.e.
the $\gamma^* \to \gamma^*$ impact factor. Its determination would open the 
way to predictions of the $\gamma^* \gamma^*$ total cross section and would 
represent a necessary ingredient in the study of the $\gamma^* p$ total cross 
section, relevant for deep inelastic scattering. Its calculation is 
rather complicated and only after year-long efforts it is approaching 
completion~\cite{gammaIF}.

A considerable simplification can be gained if one considers instead the impact
factor for the transition from a virtual photon $\gamma^*$ to a light 
neutral vector meson $V=\rho^0, \omega, \phi$. In this case, indeed, a close 
analytical expression can be achieved in the NLA, up to contributions
suppressed as inverse powers of the photon virtuality~\cite{IKP04}. 
The knowledge of the $\gamma^* \to V$ impact factor in the NLA allows 
to determine completely within perturbative QCD and with NLA accuracy the amplitude 
of a physical process, the $\gamma^* \gamma^* \to V V$ reaction~\cite{EPSW05,IP05}
(Section~5). This possibility is interesting first of all for theoretical reasons, 
since it could be used as a test-ground for comparisons with approaches different 
from BFKL, such as DGLAP, and with possible next-to-leading order extensions of 
phenomenological models. Moreover, it is useful for testing the improvement methods 
suggested at the level of the NLA Green's function for solving the problem of the 
instability of the perturbative series. But it could be interesting also for possible 
applications to phenomenology. Indeed, the calculation of the 
$\gamma^* \to V$ impact factor is the first step towards the application of the
BFKL approach to the description of processes such as the vector meson 
electroproduction $\gamma^* p\to V p$, being carried out at the HERA collider, 
and the production of two mesons in the photon collision, $\gamma^*\gamma^*\to VV$ 
or $\gamma^* \gamma \to VJ/\Psi$, which can be studied at high-energy $e^+e^-$ and 
$e\gamma$ colliders.

\section{Gluon Reggeization in perturbative QCD}

The key role in the derivation of the BFKL equation is played by the gluon 
Reggeization. ``Reggeization'' of a given elementary particle usually means that
the amplitude of a scattering process with exchange of the quantum numbers 
of that particle in the $t$-channel goes like $s^{j(t)}$ in the 
Regge limit $s\gg |t|$. The function $j(t)$ is called ``Regge trajectory''
of the given particle and takes the value of the spin of that particle when 
$t$ is equal to its squared mass. In perturbative QCD, the notion of gluon 
Reggeization is used in a stronger sense. It means not only that a Reggeon 
exists with the quantum numbers of the gluon and with a trajectory $j(t) = 
1 + \omega(t)$ passing through 1 at $t=0$, but also that this Reggeon gives 
the leading contribution in each order of perturbation theory to the amplitude 
of processes with large $s$ and fixed (i.e. not growing with $s$) squared momentum 
transfer $t$.

\begin{figure}[tb]
\begin{minipage}{60mm}
\begin{center}
\includegraphics*[width=4cm]{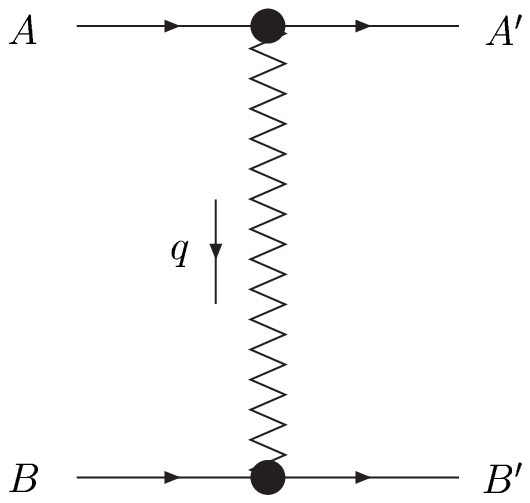}
\end{center}
\end{minipage}
\begin{minipage}{60mm}
\begin{center}
\includegraphics*[width=5.1cm]{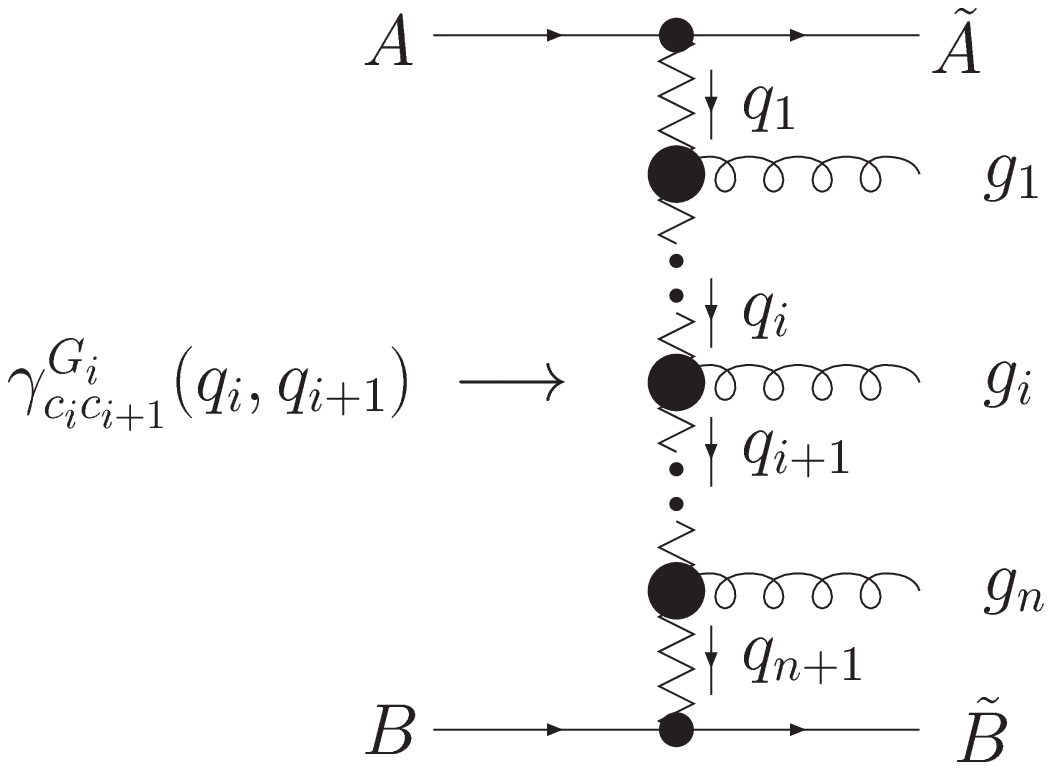}
\end{center}
\end{minipage}

\vspace{0.5cm}
\noindent {\scriptsize
Fig.~1. (Left) Diagrammatic representation of $(A_8^-)^{A^\prime B^\prime}_{AB}$.
The zig-zag line is the Reggeized gluon, the black blobs are the PPR effective 
vertices.} 

\vspace{0.3cm}
\noindent {\scriptsize
Fig.~2. (Right) Diagrammatic representation of the production amplitude
$A^{\tilde A  \tilde B +n}_{AB}$ in the LLA.}
\vspace{0.5cm}

\end{figure}

To be definite, let us consider the elastic process $A + B \longrightarrow A^\prime 
+ B^\prime $ with exchange of gluon quantum numbers in the $t$-channel, i.e.
for octet color representation in the $t$-channel and negative 
signature~\footnote{The {\em negative (positive) signature} part of an amplitude is 
the part of the amplitude which is {\em odd (even)} under the exchange of the 
Mandelstam variables $s$ and $u$.} (see Fig.~1). Gluon Reggeization means that, in 
the Regge kinematical region $s \simeq - u \rightarrow \infty$, $t$ fixed (i.e. not 
growing with $s$), the amplitude of this process takes the form
\begin{equation}
\left({A_8^-}\right)^{A^\prime B^\prime}_{AB} = 
\Gamma^c_{A^\prime A}\:\left[\left({-s\over -t}\right)^{j(t)}-\left({s\over 
-t}\right)^{j(t)}\right]\:\Gamma^c_{B^\prime B}\;.
\label{elast_ampl_octet}
\end{equation}
Here $c$ is a color index and $\Gamma^c_{P^\prime P}$ are the 
particle-particle-Reggeon (PPR) vertices, not depending on $s$. This form of the 
amplitude has been proved rigorously~\cite{BFL79} to all orders of perturbation 
theory in the LLA.
In this approximation the Reggeized gluon trajectory, $j(t)\equiv 1+\omega(t)$, enters 
with 1-loop accuracy~\cite{Lip76}, and we have
$$
\omega^{(1)}(t) = {g^2 t\over {(2{\pi})}^{D-1}}\frac{N}{2}
\int{d^{D-2}k_\perp\over k_\perp^2{(q-k)}_{\perp}^2}
=-\frac{g^2 N \Gamma(1-\epsilon)}{(4\pi)^{D/2}} 
\frac{\Gamma^2(\epsilon)}{\Gamma(2\epsilon)}(-q_\perp^2)^\epsilon\;.
$$
Here $D=4+2 \epsilon$ has been introduced in order to regularize the infrared 
divergences and the integration is performed in the space transverse to the 
momenta of the initial colliding particles~\footnote{In the following, since the 
transverse component of any momentum is obviously space-like, the notation 
$p_\perp^2 = - \vec {p}^{\:2}$ will be also used.}. 
In the NLA the form~(\ref{elast_ampl_octet}) has been checked in the first three 
orders of perturbation theory~\cite{FFKQ95-96} and is only assumed to be valid to 
all orders. 

\begin{figure}[tb]
\begin{minipage}{40mm}
\includegraphics*[width=4cm]{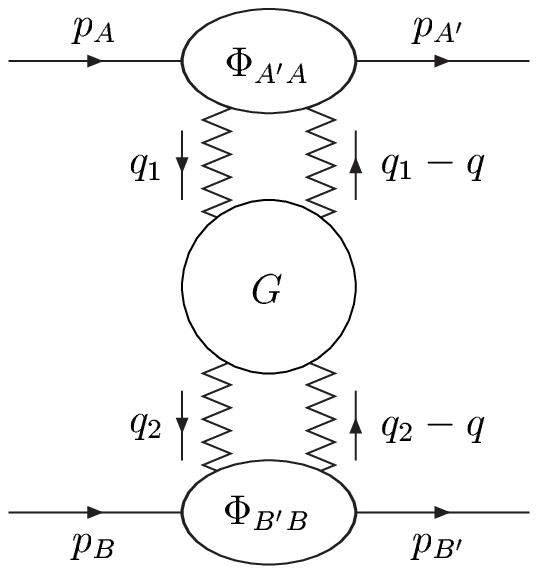}
\end{minipage}
\begin{minipage}{95mm}
\begin{flushright}
\includegraphics*[width=8cm]{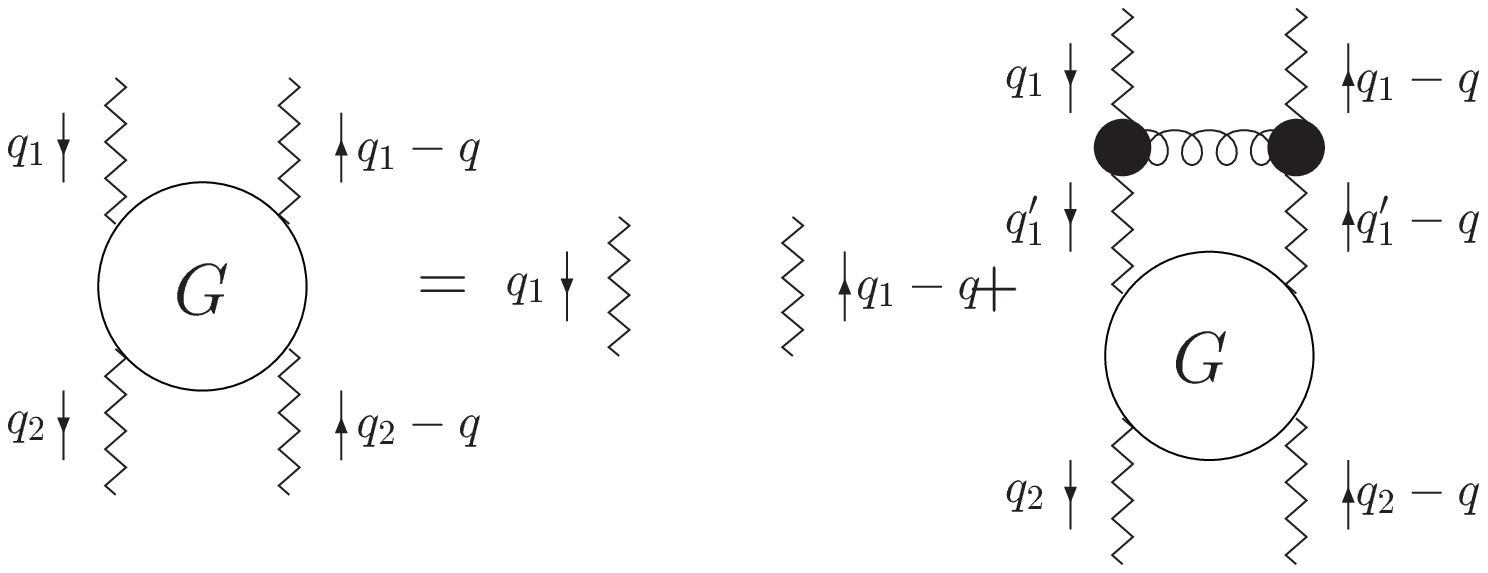}
\end{flushright}
\end{minipage}

\vspace{0.5cm}
\noindent {\scriptsize
Fig.~3. (Left) Diagrammatic representation of $A_{AB}^{A^{\prime }B^{\prime }}$ 
(for a definite color group representation) as derived from $s$-channel unitarity. 
The ovals are the impact factors of the particles $A$ and $B$, the circle is 
the Green's function for the Reggeon-Reggeon scattering.}

\vspace{0.3cm}
\noindent {\scriptsize
Fig.~4. (Right) Schematic representation of the BFKL equation in the LLA.}
\vspace{0.5cm}

\end{figure}

\section{BFKL in the LLA and the ``bootstrap''}

Amplitudes with quantum numbers in the $t$-channel different from the gluon ones
are obtained in the BFKL approach by means of unitarity relations, thus calling 
for inelastic amplitudes. In the LLA, the main contributions to the unitarity 
relations from inelastic amplitudes come from the multi-Regge kinematics, i.e. 
when rapidities of the produced particles are strongly ordered and their transverse 
momenta do not grow with $s$. In the multi-Regge kinematics, the real 
part~\footnote{The imaginary part gives a next-to-next-to-leading contribution in 
the unitarity relations.} of the production amplitudes takes a simple factorized
form, due to gluon Reggeization,
\begin{equation}
A^{\tilde A  \tilde B +n}_{AB}=2s \Gamma_{\tilde A A}^{c_1} \!
\left(\prod_{i=1}^n \gamma _{c_ic_{i+1}}^{P_i}(q_i,q_{i+1})
\left(\frac{s_i}{s_R}\right) ^{\omega_i}\! \frac 1{t_i}\right) \!
\frac 1{t_{n+1}}\!
\left( \frac{s_{n+1}}{s_R}\right) ^{\omega_{n+1}}\!
\Gamma _{\tilde B B}^{c_{n+1}}\,,
\label{inelast_ampl}
\end{equation}
where $s_R$ is an energy scale, irrelevant in the LLA, 
$\gamma_{c_i c_{i+1}}^{P_i}(q_i,q_{i+1})$ is the (non-local) effective vertex 
for the production of the particles $P_i$ with momenta $k_i=q_i-q_{i+1}$ 
in the collisions of Reggeons with momenta $q_i$ and $-q_{i+1}$ and color indices
$c_i$ and $c_{i+1}$, $q_0\equiv p_A$, $q_{n+1}\equiv -p_B$, $s_i=(k_{i-1}+k_i)^2$,
$k_0 \equiv p_{\tilde A}$, $k_{n+1} \equiv p_{\tilde B}$ and $\omega_i$ stands 
for $\omega(t_i)$, with $t_i=q_i^2$. In the LLA, $P_i$ can 
be only the state of a single gluon (see Fig.~2). By using $s$-channel unitarity 
and the previous expression for the production amplitudes, the amplitude of the
elastic scattering process $A + B \longrightarrow A^\prime + B^\prime $ at high
energies can be written as
\begin{eqnarray}
A_{AB}^{A^{\prime }B^{\prime }} &=& \frac{is}{{\left( 2\pi \right) ^{D-1}}}
\int\frac{d^{D-2}q_{1}}{\vec{q}_{1}^{~2} \vec{q}_{1}^{\, \prime\, 2}}\int
\frac{d^{D-2}q_{2}}{\vec{q}_{2}^{~2} \vec{q}_{2}^{\, \prime \, 2}}
\int_{\delta-i\infty}^{\delta+i\infty} \frac{d\omega}{\mbox{sin}(\pi\omega)}
\sum_{R,\nu} \Phi _{A^{\prime}A}^{\left(R,\nu \right) }\left(
\vec{q}_{1};\vec{q};s_{0}\right)
\nonumber \\
&\times& \left[\left( \frac{-s}{s_{0}}\right)^{\omega}-
\tau\left(\frac{s}{s_{0}}\right)^{\omega}\right]
{G_{\omega }^{\left(R\right) }\left( \vec{q}_{1},\vec{q}_{2},\vec{q}\right)}
\Phi_{B^{\prime }B}^{\left(R,\nu \right)}\left(-\vec{q}_{2};-\vec{q};s_{0}\right)\,.
\label{elast_ampl_R}
\end{eqnarray}
Here and below $q_i^\prime \equiv q_i - q$, $q\sim q_\perp$ is the momentum 
transfer in the process, the sum is over the irreducible representations $R$
of the color group obtained in the product of two
adjoint representations and over the states $\nu$ of these representations, 
$\tau$ is the signature equal to $+1 (-1)$ for symmetric (antisymmetric) 
representations and $s_{0}$ is an artificial energy scale, which disappears in the 
full expression of the amplitude at each fixed order of approximation. 
$\Phi_{A^{\prime }A}^{\left(R,\nu \right)}$ and $\Phi_{B^{\prime }B}^{\left(R,\nu \right)}$
are the so-called impact factors in the $t$-channel color state $(R,\nu)$. The first of 
them is related to the convolution of the PPR effective vertices $\Gamma_{\tilde A A}$ and
$\Gamma_{A^\prime \tilde A}$, the second to the convolution of $\Gamma_{\tilde B B}$
and $\Gamma_{B^\prime \tilde B}$~\footnote{For the precise definition of impact factors,
as well as of the BFKL kernel $K^{(R)}$ which appears below, see Ref.~\cite{FF98}.}. 
$G_\omega^{(R)}$ is the Mellin transform of the Green's functions 
for Reggeon-Reggeon scattering (see Fig.~3). The dependence from $s$ is determined by 
$G_\omega^{(R)}$, which obeys the equation (see Fig.~4)
\begin{eqnarray}
\omega G_{\omega }^{(R)}(\vec{q}_{1},\vec{q}_{2},\vec{q}) &=& 
\vec{q}_{1}^{\:2}\vec{q}_{1}^{\:\prime\:2}
\delta^{(D-2)}(\vec{q}_{1}-\vec{q}_2) \nonumber \\
&+&\int \frac{d^{D-2}{q}_r}{\vec{q}_r^{\:2}(\vec{q}_r-\vec q)^2}
K^{(R)}(\vec{q}_{1},\vec{q}_r;\vec{q}) G_{\omega}^{(R)}(\vec{q}_r,\vec{q}_{2};
\vec{q})\;,
\label{BFKL}
\end{eqnarray}
whose integral kernel, 
\begin{equation}
K^{(R)}(\vec{q}_{1},\vec{q}_{2};\vec{q}) 
=[ \omega(-\vec{q}_{1}^{\:2})+\omega(-\vec{q}_{1}^{\:\prime\:2})]
\delta^{(D-2)}(\vec{q}_{1}-\vec{q}_{2}) 
+K_{r}^{(R)}(\vec{q}_{1},\vec{q}_{2};\vec{q})\;,
\label{kernel}
\end{equation}
is composed by a ``virtual'' part, related to the gluon trajectory, 
and by a ``real'' part, $K_r^{(R)}$, related to particle production in Reggeon-Reggeon 
collisions. In the LLA, the ``virtual'' part of the kernel takes contribution
from the gluon Regge trajectory with 1-loop accuracy, $\omega^{(1)}$, while
the ``real'' part takes contribution from the production
of one gluon in the Reggeon-Reggeon collision at Born level, 
$K_{RRG}^{(B)}$. The BFKL equation is given by Eq.~(\ref{BFKL}) when $t=0$ and for 
singlet color representation in the $t$-channel, otherwise Eq.~(\ref{BFKL}) is called the
``generalized'' BFKL equation. 

The representation~(\ref{elast_ampl_R}) of the elastic amplitude, 
$A + B \longrightarrow A^\prime + B^\prime$, derived from $s$-channel unitarity,
for the part with gluon quantum numbers in the $t$-channel ($R=8$, $\tau=-1$),
must reproduce the representation~(\ref{elast_ampl_octet}) with one Reggeized 
gluon exchange in the $t$-channel, with LLA accuracy. This consistency is called
``bootstrap'' and was checked in the LLA already in Ref.~\cite{BFKL}. Subsequently,
a rigorous proof of the gluon Reggeization in the LLA was constructed~\cite{BFL79}.

The part of the representation~(\ref{elast_ampl_R}) with vacuum quantum numbers
in the $t$-channel ($R=0$, $\tau=+1$) for the case of zero momentum transfer
is relevant for the total cross section of the scattering of particles $A$ and $B$,
via the optical theorem:
\beq{sigmatot}
\sigma_{AB}(s)=
\frac{{\cal I}m_s {\cal A}_{AB}^{AB}}{s}=
\int\frac{d^2\vec q_1}{2\pi}\frac{\Phi_A( \vec q_1,s_0)}{\vec q_1^{\,\,2}}
\int\frac{d^2\vec q_2}{2\pi}\frac{\Phi_B(-\vec q_2,s_0)}{\vec q_2^{\,\,2}} 
\eeq
\[
\times
\int\limits^{\delta +i\infty}_{\delta -i\infty}\frac{d\omega}{2\pi i}
\left(\frac{s}{s_0}\right)^\omega G_\omega (\vec q_1, \vec q_2)\;,
\]
where $D$ has been put equal to 4, assuming that the state $A$ and $B$ are
colorless and have therefore good infrared behavior~\cite{FM99}, and
\[
G_\omega (\vec q_1, \vec q_2) \equiv \frac{G_\omega^{(0)} (\vec q_1, \vec q_2)}
{\vec q_1^{\:2} \vec q_2^{\:2}}\;.
\]
This Green's function satisfies the equation
\beq{Green}
\omega \, G_\omega (\vec q_1, \vec q_2)=\delta^2(\vec q_1-\vec q_2)+
\int d^2 \vec q \, K(\vec q_1,\vec q)\, G_\omega (\vec q, \vec q_2) \;,
\eeq
where
\beq{ker}
K(\vec q_1, \vec q_2) \equiv \frac{K^{(0)} (\vec q_1, \vec q_2)}
{\vec q_1^{\:2} \vec q_2^{\:2}}\;.
\eeq
The eigenfunctions and the corresponding eigenvalues of the LLA singlet kernel
are known (see, for instance, Ref.~\cite{news}):
\[
\int d^2 \vec q_2 K(\vec q_1, \vec q_2) (\vec q_2^{\:2})^{\gamma-1}=
\frac{N_c \alpha_s}{\pi} \chi(\gamma) (\vec q_1^{\:2})^{\gamma-1}\;,
\]
with
\[
\chi(\gamma)=2 \psi(1)-\psi(\gamma)-\psi(1-\gamma)\;,
\hspace{2cm} \psi(\gamma)\equiv \frac{\Gamma'(\gamma)}{\Gamma(\gamma)}\;.
\]
The set of functions $(\vec q_2^{\:2})^{\gamma-1}$ with $\gamma=1/2+i \nu$, 
$-\infty < \nu < \infty$ forms a complete set, so that 
\beq{sigmanu}
\sigma_{AB}(s)= 
\int\limits^{\delta +i\infty}_{\delta -i\infty}\frac{d\omega}{2\pi i}
\int_{-\infty}^{+\infty} \frac{d\nu}{2\pi^2 [\omega-\frac{N_c\alpha_s}{\pi}
\chi(1/2+i\nu)]}
\eeq
\[
\times 
\int \frac{d^2\vec q_1}{2\pi}\Phi_A(\vec q_1,s_0)
\int \frac{d^2\vec q_2}{2\pi}\Phi_B(-\vec q_2,s_0)
\left(\frac{s}{s_0}\right)^\omega 
(\vec q_1^{\:2})^{-i\nu-3/2}
(\vec q_2^{\:2})^{+i\nu-3/2}\;.
\]
The maximal value of $\chi(\gamma)$ on the integration contour is $\chi(1/2)=4 \ln 2$,
which corresponds to the maximal eigenvalue of the kernel 
$\omega_P^B=(\alpha_s N_c/\pi) 4 \ln 2$, where the subscript ``$P$'' stands
for ``Pomeron''. It is then easy to see that 
Eq.~(\ref{sigmanu}) leads, by saddle point evaluation of the $\nu$-integration, to
the following result:
\beq{sigmaLLA}
\sigma^{\mbox{\tiny LLA}} \sim \frac{s^{\omega_P^B}}{\sqrt{\ln s}}\;.
\eeq
For $\alpha_s=0.15$ one gets $\omega_P^B\simeq 0.40$, much larger than the 
corresponding value for the cross section of soft hadronic processes ($\simeq$ 0.08), 
but in rough agreement with the power-growth in $x$ observed for the gluon density
in the proton at small $x$ and large virtuality $Q^2$.

The relation~(\ref{sigmaLLA}) shows that unitarity is violated, since 
the cross section overcomes the Froissart-Martin bound. This is obvious since,
in the LLA, only a definite set of intermediate states, as we have seen, contributes
to the $s$-channel unitarity relation. This means that the BFKL approach cannot be 
applied at asymptotically high energies. In order to identify the applicability
region of the BFKL approach, it is necessary to know the scale of 
$s$ and the argument of the running coupling constant, which are not
fixed in the LLA.

\section{BFKL in the NLA}

In the NLA, the Regge form of the elastic amplitude~(\ref{elast_ampl_octet})
and of the production amplitudes~(\ref{inelast_ampl}), implied by gluon
Reggeization, has been checked only in the first three orders of perturbation 
theory~\cite{FFKQ95-96}. In order to derive the BFKL equation in the NLA, gluon
Reggeization is assumed to be valid to all orders of perturbation theory. It becomes
important, therefore, to check the validity of this assumption. 

In the NLA it is necessary to include into the unitarity relations contributions
which differ from those in the LLA by having one additional power of $\alpha_s$
or one power less in $\ln s$. The first set of corrections is realized by
performing, only in one place, one of the following replacements in the production 
amplitudes~(\ref{inelast_ampl}) entering the $s$-channel 
unitarity relation:
$$
\omega^{(1)} \longrightarrow \omega^{(2)}\;,
\;\;\;\;\;\;\;\;\;\;
\Gamma_{P'P}^{c\;\mbox{\scriptsize (Born)}} \longrightarrow 
\Gamma_{P'P}^{c\;\mbox{\scriptsize (1-loop)}}\;,
\;\;\;\;\;\;\;\;\;\;
\gamma_{c_i c_{i+1}}^{G_i \mbox{\scriptsize (Born)}} \longrightarrow 
\gamma_{c_i c_{i+1}}^{G_i \mbox{\scriptsize (1-loop)}}\;.
$$
The second set of corrections consists in allowing the production in the 
$s$-channel intermediate state of {\em one} pair of particles with rapidities 
of the same order of magnitude, both in the central or in the fragmentation 
region (quasi-multi-Regge kinematics). This implies one replacement among the 
following ones in the production amplitudes~(\ref{inelast_ampl}) entering the 
$s$-channel unitarity relation: 
$$
\Gamma_{P'P}^{c\;\mbox{\scriptsize (Born)}} \longrightarrow 
\Gamma_{\{f\} P}^{c\;\mbox{\scriptsize (Born)}}\;,
\;\;\;\;\;\;\;\;\;\;\;\;\;\;\;\;\;\;\;\;\;\;
\gamma_{c_i c_{i+1}}^{G_i \mbox{\scriptsize (Born)}} \longrightarrow 
\gamma_{c_i c_{i+1}}^{Q\overline Q \mbox{\scriptsize (Born)}}\;,
$$
$$
\gamma_{c_i c_{i+1}}^{G_i \mbox{\scriptsize (Born)}} \longrightarrow 
\gamma_{c_i c_{i+1}}^{GG \mbox{\scriptsize (Born)}}\;.
$$
Here $\Gamma_{\{f\} P}$ stands for the production of a state containing an extra
particle in the fragmentation region of the particle $P$ in the scattering
off the Reggeon, $\gamma_{c_i c_{i+1}}^{Q\overline Q \mbox{\scriptsize (Born)}}$
and $\gamma_{c_i c_{i+1}}^{GG \mbox{\scriptsize (Born)}}$ are
the effective vertices for the production of a quark anti-quark pair and
of a two-gluon pair, respectively, in the collision of two Reggeons. 

The detailed program of next-to-leading corrections to the BFKL equation
was formulated in Ref.~\cite{FL89} and was carried out over a period
of several years (for an exhaustive review, see Ref.~\cite{news}). It turns out 
that also in the NLA the amplitude for the high energy elastic process 
$A + B \longrightarrow A^\prime + B^\prime$ can be represented as in 
Eq.~(\ref{elast_ampl_R}) and in Fig.~3. The Green's function
obeys an equation with the same form as Eq.~(\ref{BFKL}), with a kernel having
the same structure as in Eq.~(\ref{kernel}). Here the ``virtual'' part of the kernel
takes also the contribution from the gluon trajectory at 2-loop accuracy, 
$\omega^{(2)}$~\cite{FFKQ95-96}, while the ``real'' part of the kernel 
takes the additional contribution from one-gluon production in the Reggeon-Reggeon 
collisions at 1-loop order, $K_{RRG}^{(1)}$~\cite{FL93,FFQ94,FFK96,FFP01}, 
from quark anti-quark pair production at Born level, 
$K_{RRQ\overline Q}^{(B)}$ (Ref.~\cite{FFFK97} for the forward case, 
Ref.~\cite{FFP99} for the non-forward case) and from two-gluon pair production at 
Born level, $K_{RRGG}^{(B)}$ (Ref.~\cite{FKL97} for the forward case,
Ref.~\cite{FG00} for the non-forward, octet case, Ref.~\cite{FF05} for the 
non-forward, singlet case). 

The consistency between the representation~(\ref{elast_ampl_R}) 
of the elastic amplitude, $A + B \longrightarrow A^\prime + B^\prime$, derived 
from $s$-channel unitarity, for the part with gluon quantum numbers in the 
$t$-channel ($R=8$, $\tau=-1$), and the representation~(\ref{elast_ampl_octet}) 
with one Reggeized gluon exchange in the $t$-channel (``bootstrap'') is of crucial
importance in the NLA. In this approximation, indeed, gluon Reggeization was only 
assumed in order to derive the BFKL equation. Moreover, the check of the 
bootstrap is also a (partial) check of the correctness of calculations
which were performed mostly by one research group.
In the NLA, the bootstrap leads to two conditions to be fulfilled~\cite{FF98},
one on the NLA octet kernel, the other on the NLA octet impact factors,
which have both been verified~\cite{FFP99,FFK00,FFKP00}. 

Another set of bootstrap conditions, proposed in Ref.~\cite{BV99}, was
introduced. They were called ``strong'' bootstrap conditions, since 
their fulfillment implies that of the bootstrap conditions considered so far. They
constrain the form of the octet impact factors and of the octet BFKL kernel.
Their fulfillment has been verified in Refs.~\cite{FFKP00_strong,FP02}.
Recently is has been understood that these strong bootstrap conditions
are a subset of those which arise from the request of consistency 
with $s$-channel unitarity of the inelastic amplitudes in the Regge form
which enter the BFKL approach~\cite{Fad05}. Their fulfillment amounts to
prove the gluon Reggeization in the NLA~\cite{Fad05}. 

As for the exchange of vacuum quantum numbers in the $t$-channel, the NLA
corrections to the BFKL kernel lead to a large correction to the BFKL Po\-me\-ron 
intercept~\cite{FL98,CC98}. Indeed, one gets now~\cite{news} 
\[
\int d^2 \vec q_2 K(\vec q_1, \vec q_2) (\vec q_2^{\:2})^{\gamma-1}=
\frac{N_c \alpha_s(\vec q_1^{\:2})}{\pi} 
\left(\chi(\gamma)+\frac{\alpha_s(\vec q_1^{\:2}) N_c}{\pi}\chi^{(1)}(\gamma)\right)
(\vec q_1^{\:2})^{\gamma-1}\;,
\]
where $\chi^{(1)}(\gamma)$ is a known function~\cite{news}. We observe that now
the argument of $\alpha_s$ is not undefined as in the LLA case. The relative
correction $r(\gamma)$ defined as $\chi^{(1)}(\gamma)=-r(\gamma) \chi(\gamma)$
in the point $\gamma=1/2+i\nu$ turns out to be very large~\cite{FL98,CC98}, 
\[
r\left(\frac{1}{2}\right)\simeq 6.46 +0.05 \frac{n_f}{N_c}+0.96 \frac{n_f}{N_c^3}
\]
and leads to a NLA Pomeron intercept
\[
\omega_P=\omega_P^B (1- 2.4 \: \omega_P^B)\;, \hspace{1cm}
\omega_P^B = 4 \ln 2 N \alpha_s(\vec q^{\:2})/\pi\;.
\]
A lot of papers have been devoted to the problem of this large correction (see, for 
instance,~\cite{NLA_Pomeron}). As anticipated in the Introduction, the understanding of the 
way to treat these large corrections can be helped by considering the full NLA amplitude 
of hard QCD physical processes, instead of limiting the attention to the unphysical 
BFKL Green's function. The construction of a physical NLA amplitude in the BFKL approach,
however, calls for the determination of the NLA impact factors of the colorless 
particles involved in the process. This determination can be achieved by means of 
perturbation theory only in case a hard enough scale exists, such as a large photon 
virtuality or the mass of a heavy quark.

\section{The inclusion of impact factors: the amplitude for the electroproduction
of two light vector mesons}

Recently, the impact factor for the transition from a virtual photon with
longitudinal polarization to a light vector meson with longitudinal
polarization was calculated with NLA accuracy~\cite{IKP04}. 
It can be used, together with the NLA BFKL Green's function to build with
NLA accuracy the amplitude for the production of two light vector mesons 
($V=\rho^0, \omega, \phi$) in the collision of two virtual photons.
In the kinematics $s\gg Q^2_{1,2}\gg \Lambda^2_{QCD}$, where $Q^2_{1,2}$ are
the photon virtualities, other helicity amplitudes are indeed power suppressed, 
with a suppression factor $\sim m_V/Q_{1,2}$ and the light vector meson mass can be put
equal to zero.

As we have seen before, the forward amplitude may be presented as follows
\beq{imA}
{\cal I}m_s\left( {\cal A} \right)=\frac{s}{(2\pi)^2}\int\frac{d^2\vec q_1}{\vec
q_1^{\,\, 2}}\Phi_1(\vec q_1,s_0)\int
\frac{d^2\vec q_2}{\vec q_2^{\,\,2}} \Phi_2(-\vec q_2,s_0)
\int\limits^{\delta +i\infty}_{\delta
-i\infty}\frac{d\omega}{2\pi i}\left(\frac{s}{s_0}\right)^\omega
G_\omega (\vec q_1, \vec q_2).
\eeq
This representation for the amplitude is valid with NLA accuracy.
Here $\Phi_{1}(\vec q_1,s_0)$ and $\Phi_{2}(-\vec q_2,s_0)$
are the impact factors describing the transitions $\gamma^*(p)\to V(p_1)$
and $\gamma^*(p')\to V(p_2)$, respectively.
The Green's function in (\ref{imA}) obeys the BFKL equation~(\ref{Green}).
The scale $s_0$ is artificial and must disappear in the full expression for the amplitude
at each fixed order of approximation~\footnote{For understanding the reasons why this
scale was introduced and enters also the definition of the impact factors, see 
Refs.~\cite{news,FF98}.}. Using the result for the meson
NLA impact factor such cancellation was demonstrated explicitly in
Ref.~\cite{IKP04} for the process in question.

The impact factors are known as an expansion in $\alpha_s$ (see Ref.~\cite{IKP04})
\beq{impE}
\Phi_{1,2}(\vec q)= \alpha_s \,
D_{1,2}\left[C^{(0)}_{1,2}(\vec q^{\,\, 2})+\bar\alpha_s
C^{(1)}_{1,2}(\vec
q^{\,\, 2})\right] \, , \quad D_{1,2}=-\frac{4\pi e_q  f_V}{N_c Q_{1,2}}
\sqrt{N_c^2-1}\, ,
\eeq
where $f_V$ is the meson dimensional coupling constant ($f_{\rho}\approx
200\, \rm{ MeV}$) and $e_q$ should be replaced by $e/\sqrt{2}$, $e/(3\sqrt{2})$
and $-e/3$ for the case of $\rho^0$, $\omega$ and $\phi$ meson production,
respectively.

Using the known impact factors and the NLA Green's function, spectrally decomposed
on the basis of the LLA kernel eigenfunctions, one gets~\cite{IP05}
\[
\frac{{\cal I}m_s\left( {\cal A} \right)}{D_1D_2}=\frac{s}{(2\pi)^2}
\int\limits^{+\infty}_{-\infty}
d\nu \left(\frac{s}{s_0}\right)^{\bar \alpha_s(\mu_R) \chi(\nu)}
\alpha_s^2(\mu_R) c_1(\nu)c_2(\nu)
\]
\beq{amplnla}
\times \left[1+\bar \alpha_s(\mu_R)\left(\frac{c^{(1)}_1(\nu)}{c_1(\nu)}
+\frac{c^{(1)}_2(\nu)}{c_2(\nu)}\right) \right.
\eeq
\[
\left.
+\bar \alpha_s^2(\mu_R)\ln\left(\frac{s}{s_0}\right)
\left(\bar
\chi(\nu)+\frac{\beta_0}{8N_c}\chi(\nu)\left[-\chi(\nu)+\frac{10}{3}
+i\frac{d\ln(\frac{c_1(\nu)}{c_2(\nu)})}{d\nu}+2\ln(\mu_R^2)\right]
\right)\right] \; ,
\]
where $c_{1,2}(\nu)$ ($c_{1,2}^{(1)}(\nu)$) are the coefficient of the expansion of 
the LLA (NLA) impact factors for the two photons in the basis formed by the 
eigenfunctions of the LLA kernel. All the other functions of $\nu$ in~(\ref{amplnla})
are known (see Ref.~\cite{IP05}).
It is possible to write this amplitude in the form of a series, as
\bear{series}
\frac{Q_1Q_2}{D_1 D_2} \frac{{\cal I}m_s {\cal A}}{s} &=&
\frac{1}{(2\pi)^2}  \alpha_s(\mu_R)^2 \label{honest_NLA} \\
& \times &
\biggl[ b_0
+\sum_{n=1}^{\infty}\bar \alpha_s(\mu_R)^n   \, b_n \,
\biggl(\ln\left(\frac{s}{s_0}\right)^n   +
d_n(s_0,\mu_R)\ln\left(\frac{s}{s_0}\right)^{n-1}     \biggr)
\biggr]\;, \nonumber
\eear
where the $b_n$ and $d_n$ coefficients can be easily determined 
by comparison with~(\ref{amplnla}).
One should stress that both representations of the amplitude~(\ref{series})
and~(\ref{amplnla}) are equivalent with NLA accuracy, since they differ only by
next-to-NLA (NNLA) terms. It is easily seen from Eq.~(\ref{series}) that the 
amplitude is independent in the NLA from the choice of energy and strong coupling
scales. Indeed, with the required accuracy,
\beq{asev}
\bar \alpha_s(\mu_R)=\bar \alpha_s(\mu_0)\left(
1-\frac{\bar \alpha_s(\mu_0)\beta_0}{4N_c}\ln\left(\frac{\mu_R^2}{\mu_0^2}\right)
\right)
\eeq
and therefore terms $\bar \alpha_s^n\ln^{n-1} s\ln s_0$ and $\bar
\alpha_s^n\ln^{n-1} s\ln \mu_R$ cancel in~(\ref{series}).

Here some numerical results are presented for the amplitude given in Eq.~(\ref{series}) 
for the $Q_1=Q_2\equiv Q$ kinematics, i.e. in the
``pure'' BFKL regime. The other interesting regime, $Q_1\gg Q_2$ or vice-versa,
where collinear effects could come heavily into the game, will not be 
considered here. In the numerical analysis presented below the series in the
R.H.S. of Eq.~(\ref{series}) has been truncated to $n=20$, after having verified that
this procedure gives a very good approximation of the infinite sum for $Y\leq 10$.

The $b_n$ and $d_n$ coefficients for $n_f=5$ and $s_0=Q^2=\mu_R^2$ can be calculated
numerically and turn to be the following ones:
\beq{coe}
\begin{array}{llll}
b_0=17.0664  &              &              &              \\ 
b_1=34.5920  & b_2=40.7609  & b_3=33.0618  & b_4=20.7467  \\
b_5=10.5698  & b_6=4.54792  & b_7=1.69128  & b_8=0.554475 \\
             &              &              &              \\
d_1=-3.71087 & d_2=-11.3057 & d_3=-23.3879 & d_4=-39.1123 \\
d_5=-59.207  & d_6=-83.0365 & d_7=-111.151 & d_8=-143.06\;. \\
\end{array}
\eeq
These numbers make visible the effect of the NLA corrections: the $d_n$
coefficients are negative and increasingly large in absolute values as the
perturbative order increases. 
In this situation the optimization of perturbative expansion, in our case the
choice of the renormalization scale $\mu_R$ and of the energy scale $s_0$,
becomes an important issue.
Below the principle of minimal sensitivity (PMS)~\cite{Stevenson} is adopted.
Usually PMS is used to fix the value of the renormalization scale for the strong
coupling. Here this principle is used in a broader sense, requiring 
the minimal sensitivity of the predictions to the change of both the renormalization 
and the energy scales, $\mu_R$ and $s_0$.
More precisely, we replace in~(\ref{series}) $\ln(s/s_0)$ with $Y-Y_0$,
where $Y=\ln(s/Q^2)$ and $Y_0=\ln(s_0/Q^2)$, and study the dependence of the
amplitude on $Y_0$.

\begin{figure}[tb]
\centering
\hspace{-1cm}
{\epsfysize 8cm \epsffile{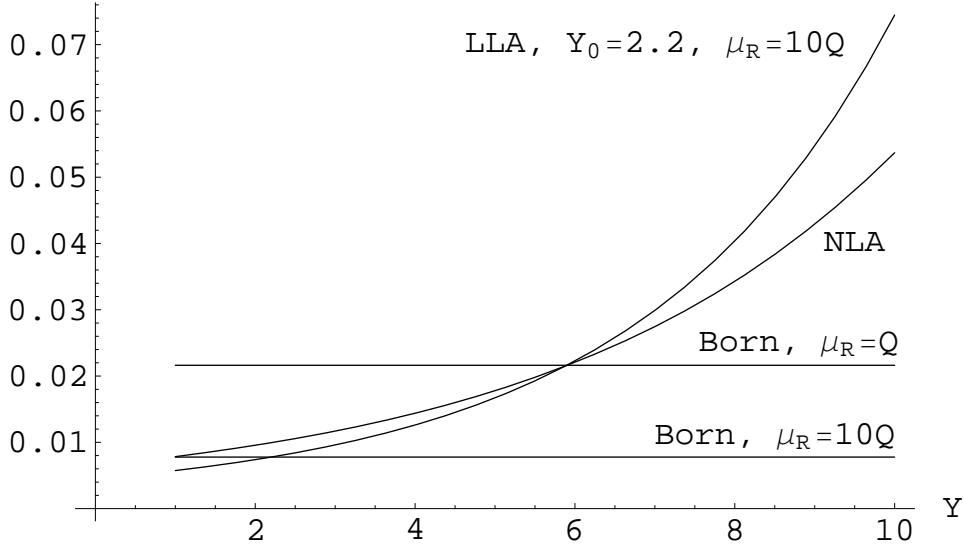}}
\caption[]{${\cal I}m_s ({\cal A})Q^2/(s \, D_1 D_2)$ as a function of
$Y$ for optimal choice of the energy parameters $Y_0$ and $\mu_R$ (curve 
labeled
by ``NLA''). The other curves represent the LLA result for $Y_0=2.2$ and $\mu_R=10Q$
and the Born (2-gluon exchange) limit for $\mu_R=Q$ and $\mu_R=10Q$.
The photon virtuality $Q^2$ has been fixed to 24 GeV$^2$ ($n_f=5$).}
\label{PMSres}
\end{figure}

It has been found that, for several fixed values of $Q^2$ and $n_f$, at any value
of the energy $Y$ there are wide regions of values of the parameters $Y_0$
and $\mu_R$ where the amplitude is practically flat, this allowing to
assign a value for the amplitude for each energy $Y$. In Fig.~\ref{PMSres} 
the behavior of the amplitude for the $Q^2$=24 GeV$^2$ and $n_f=5$ is presented.
This result shows that, although the NLA corrections are large and with opposite
sign with respect to the LLA result, it is possible to give a definite 
meaning to a perturbative series, by the request of renorm-invariance 
and of stability with respect to the artificial parameter $s_0$. 

It turns out, however, that the optimal values for $Y_0$ and $\mu_R$ are far from the 
``natural'' choices $s_0 \simeq \mu_R^2 \simeq Q^2$. In the case of 
$Q^2$=24 GeV$^2$ and $n_f=5$, we have, indeed, $Y_0\simeq 2$ and $\mu_R \simeq 10 Q$.
These values look unnatural since there is no other scale for transverse momenta 
in the problem at question except $Q$. Moreover one can guess that at higher orders
the typical transverse momenta are even smaller than $Q$ since they ``are shared"
in the many-loop integrals and the strong coupling grows in the infrared.
The large values of $\mu_R$ found is not an indication of the
appearance of a new scale, but is rather a manifestation of the nature of the
BFKL series. The fact is that NLA corrections are large and then, necessarily,
since the exact amplitude should be renorm- and energy scale invariant, the
NNLA terms should be large and of the opposite sign with respect to the NLA.
It can be guessed that if the NNLA corrections were known and if PMS were applied to the
amplitude constructed as LLA + NLA-corrections + NNLA-corrections, more ``natural'' values 
of $\mu_R$ would be obtained.

{\it It is a great pleasure for me to contribute to this volume in honor of Adriano 
Di Giacomo. His dedication to science is an example to me.}

\begin{thebibliography}{99}
\renewcommand{\itemsep}{-0.7mm}

\bibitem{BFKL} 
V.S.~Fadin, E.A.~Kuraev and L.N.~Lipatov, \Journal{\PLB}{60}{50}{1975};
E.A.~Kuraev, L.N.~Lipatov and V.S.~Fadin, \Journal{\SJETP}{44}{443}{1976};
\Journal{\SJETP}{45}{199}{1977};
Ya.Ya.~Balitskii and L.N.~Lipatov, \Journal{\SJNP}{28}{822}{1978}.

\bibitem{DGLAP} 
V.N.~Gribov and L.N.~Lipatov, \Journal{\SJNP}{15}{438}{1972};
L.N.~Lipatov, \Journal{\SJNP}{20}{94}{1975};
G.~Altarelli and G.~Parisi, \Journal{\NPB}{26}{298}{1977};
Yu.L.~Dokshitzer, \Journal{\SJETP}{46}{641}{1977}.

\bibitem{FFKP00}
V.S.~Fadin, R.~Fiore, M.I.~Kotsky and A.~Papa, \Journal{\PRD}{61}{094005}{2000};
\Journal{\PRD}{61}{094006}{2000}.

\bibitem{CR00}
M.~Ciafaloni and H.~Rodrigo, \Journal{JHEP}{0005}{042}{2000}.

\bibitem{gammaIF} 
J.~Bartels, S.~Gieseke and C.~F.~Qiao, \Journal{\PRD}{63}{056014}{2001}
[{\it Erratum} \Journal{\PRD}{65}{079902}{2002}];
J.~Bartels, S.~Gieseke and A.~Kyrieleis, \Journal{\PRD}{65}{014006}{2002};
J.~Bartels, D.~Colferai, S.~Gieseke and A.~Kyrieleis, \Journal{\PRD}{66}{094017}
{2002};
J.~Bartels and A.~Kyrieleis, \Journal{\PRD}{70}{114003}{2004};
V.S.~Fadin, D.Yu.~Ivanov and M.I.~Kotsky, \Journal{\PAN}{65}{1513}{2002},
\Journal{\NPB}{658}{156}{2003}.

\bibitem{IKP04} 
D.~Yu. Ivanov, M.I.~Kotsky and A. Papa, \Journal{\EPJC}{38}{195}{2004}; 
\Journal{\NPPS}{146}{117}{2005}.

\bibitem{EPSW05}
R.~Enberg. B.~Pire, L.~Szymanowski and S.~Wallon, hep-ph/0508134.

\bibitem{IP05}
D.~Yu. Ivanov and A. Papa, \Journal{\NPB}{732}{183}{2006}.

\bibitem{BFL79}
Ya.Ya.~Balitskii, L.N.~Lipatov and V.S.~Fadin, in {\em Proceedings of 
Leningrad Winter School on Physics of Elementary Particles} (Leningrad, 1979, p. 109).

\bibitem{Lip76}
L.N.~Lipatov, \Journal{\SJNP}{23}{338}{1976}.

\bibitem{FFKQ95-96}
V.S.~Fadin, \Journal{JETPL}{61}{346}{1995}; 
V.S.~Fadin, R.~Fiore and A.~Quartarolo, \Journal{\PRD}{53}{2729}{1996}; 
M.I.~Kotsky and V.S.~Fadin, \Journal{PAN}{59}{1035}{1996}; 
V.S.~Fadin, R.~Fiore and M.I.~Kotsky, \Journal{\PLB}{359}{181}{1995},
\Journal{\PLB}{387}{593}{1996}.

\bibitem{FF98}
V.S.~Fadin and R.~Fiore, \Journal{\PLB}{440}{359}{1998}.

\bibitem{FM99}
V.S.~Fadin and A.D.~Martin, \Journal{\PRD}{60}{114008}{1999}.

\bibitem{news}
V.S.~Fadin, hep-ph/9807528.

\bibitem{FL89}
L.N.~Lipatov and V.S.~Fadin, \Journal{\JETPL}{49}{352}{1989},
\Journal{\SJNP}{50}{712}{1989}.

\bibitem{FL93}
V.S.~Fadin and L.N.~Lipatov, \Journal{\NPB}{406}{259}{1993}.

\bibitem{FFQ94}
V.S.~Fadin, R.~Fiore and A,~Quartarolo, \Journal{\PRD}{50}{5893}{1994}.

\bibitem{FFK96}
V.S.~Fadin, R.~Fiore and M.I.~Kotsky, \Journal{\PLB}{389}{737}{1996}.

\bibitem{FFP01}
V.S.~Fadin, R.~Fiore and A.~Papa, \Journal{\PRD}{63}{034001}{2001}.

\bibitem{FFFK97}
V.S.~Fadin, R.~Fiore, A.~Flachi and M.I.~Kotsky, \Journal{\PLB}{422}{287}{1998}.

\bibitem{FFP99}
V.S.~Fadin, R.~Fiore and A.~Papa, \Journal{\PRD}{60}{074025}{1999}.

\bibitem{FKL97}
V.S.~Fadin, M.I.~Kotsky and L.N.~Lipatov, \Journal{\PLB}{415}{97}{1997}.

\bibitem{FG00}
V.S.~Fadin and D.A.~Gorbachev, \Journal{\JETPL}{71}{222}{2000};  
\Journal{PAN}{63}{2157}{2000}.

\bibitem{FF05}
V.S.~Fadin and R.~Fiore, \Journal{\PLB}{610}{61}{2005}, {\it Erratum-ibid.} {\bf 621}, 61
(2005); \Journal{PRD}{72}{014018}{2005}.

\bibitem{FFK00}
V.S.~Fadin, R.~Fiore and M.I.~Kotsky, \Journal{\PLB}{494}{100}{2000}.

\bibitem{BV99}
M.A.~Braun, hep-ph/9901447;
M.A.~Braun, G.P.~Vacca, \Journal{\PLB}{477}{156}{2000}.

\bibitem{FFKP00_strong}
V.S.~Fadin, R.~Fiore, M.I.~Kotsky and A.~Papa, \Journal{\PLB}{495}{329}{2000}.

\bibitem{FP02}
V.S.~Fadin and A.~Papa, \Journal{\NPB}{640}{309}{2002}.

\bibitem{Fad05}
V.S.~Fadin, hep-ph/0511121 and references therein.

\bibitem{FL98}
V.S.~Fadin and L.N.~Lipatov, \Journal{\PLB}{429}{127}{1998}.

\bibitem{CC98}
M.~Ciafaloni and G.~Camici, \Journal{\PLB}{430}{349}{1998}.

\bibitem{NLA_Pomeron}
D.A.~Ross, \Journal{\PLB}{431}{161}{1998};
Yu.V.~Kovchegov and A.H.~Mueller, \Journal{\PLB}{439}{423}{1998}; 
J.~Bl\"umlein, V.~Ravindran, W.L.~van Neerven and A.~Vogt, hep-ph/9806368; 
E.M.~Levin, hep-ph/9806228; 
N.~Armesto, J.~Bartels, M.A.~Braun, \Journal{\PLB}{442}{459}{1998}; 
G.P.~Salam, \Journal{\JHEP}{9807}{19}{1998}; 
M.~Ciafaloni and D.~Colferai, \Journal{\PLB}{452}{372}{1999}; 
M.~Ciafaloni, D.~Colferai and G.P.~Salam, \Journal{\PRD}{60}{114036}{1999}; 
R.S.~Thorne, \Journal{\PRD}{60}{054031}{1999};
S.J.~Brodsky, V.S.~Fadin, V.T.~Kim, L.N.~Lipatov, G.B.~Pivovarov, \Journal{\JETPL}
{70}{155}{1999};
A.~Sabio Vera, \Journal{\NPB}{722}{65}{2005}.

\bibitem{Stevenson}
P.M.~Stevenson, \Journal{\PLB}{100}{61}{1981}; \Journal{\PRD}{23}{2916}{1981}.

\end{thebibliography}
\end{document}